\documentstyle{article}

\begin{document}

\begin{center}
{\bf Infrared Dynamics in Vector-Like Gauge Theories:  QCD and Beyond}\\
\bigskip
{Victor Elias}\\
\smallskip
{\small\it Department of Applied Mathematics, The University of Western Ontario\\
London, Ontario  N6A 5B7  Canada}
\end{center}
\bigskip
\begin{abstract}
Pad\'e-approximant methods are used to extract information about 
leading positive zeros or poles of QCD and SQCD $\beta$-functions from the known 
terms of their perturbative series.  For QCD, such methods are seen to corroborate 
the flavour-threshold behaviour obtained via lattice approaches for the 
occurrence of infrared-stable fixed points.  All possible Pad\'e-approximant 
versions of the known (one- to four-loop) terms of the QCD $\overline{MS}$ 
$\beta$-function  series 
are consistent with this threshold occurring at or above $n_f=6$.  This 
conclusion continues to be true even if higher-degree Pad\'e-approximants 
are introduced to accommodate an {\it arbitrary} five-loop contribution to the 
QCD $\beta$-function for a given number of flavours.
\end{abstract}  

\bigskip
  
The dynamics characterising the infrared region of asymptotically-free gauge theories 
are becoming an area of active investigation. Historically speaking, the property of 
asymptotic freedom has been linked with ``infrared slavery'' at large distances -- the 
non-observability of gauge-group nonsinglet particles as asymptotic free-particle 
states because of the growth of the gauge coupling constant to arbitrarily large 
values at some infrared momentum scale. Such methodological simplification, however, 
rests upon the existence of a Landau singularity in the evolution of the gauge 
couplant, a singularity that is likely an artefact of truncating the gauge 
couplant's $\beta$-function to any given order. Consider, for example, the QCD 
$\beta$-function to two-loop order for three active flavours:
\begin{equation}
\mu^2 \frac{dx}{d\mu^2} = -\frac{9}{4} x^2 - 4x^3, \; \; x \equiv \alpha_s (\mu)/\pi.
\end{equation}
Given some empirical initial value, e.g. $x_\tau =  \alpha_s(m_\tau)/\pi$ , eq.(1) has the analytic solution
\begin{equation}
\log \left( \mu^2 / m_\tau^2 \right) = \frac{4}{9} \left[ \frac{1}{x(\mu)}-\frac{1}{x_\tau}\right]
+ \frac{64}{81} \log \left[ \frac{x(\mu)(x_\tau + 9/16)}{x_\tau (x(\mu)+9/16)} \right].
\end{equation}
As $\mu$ decreases from $m_\tau$ in this expression, the couplant $x(\mu)$ grows from $x_\tau$
to become infinite at
\begin{equation}
\mu_L = m_\tau \left( 1 + 9/(16x_\tau) \right)^{32/81} \exp \left[ -2/(9x_\tau)\right]
\end{equation}
the Landau singularity. The temptation is to identify $\mu_L$ with the infrared boundary of 
QCD as a perturbative gauge theory of quarks and gluons. Note that if $\alpha_s(m_\tau) = \pi
 x_\tau = 0.35$, 
the central PDG value [1], then $\mu_L$ is found from (3) to be 493 MeV.

However, this Landau singularity is really a consequence of truncating the 
perturbative  $\beta$-function series (1) to its renormalization scheme independent two-loop 
order terms. Subsequent  $\beta$-function terms are renormalization-scheme-dependent and 
presumably negotiable (I will have more to say on this later on). For example, we can 
conjecture two toy-model $\beta$-functions
\begin{equation}
\mu^2 \frac{dx}{d\mu^2} = - \frac{9}{4} \frac{x^2}{(1-16x/9)}
\end{equation}
\begin{equation}
\mu^2 \frac{dx}{d\mu^2} = - \frac{9}{4} x^2 \left( 1 + \frac{16}{9} x +R_2 x^2 \right), \; \; R_2<0
\end{equation}
which both agree to two-loop order with eq. (2), but have infrared behaviour manifestly 
different from the Landau pole characterising (2). The former equation has a $\beta$-function 
pole at $x = 9/16 = 0.5625$. Consequently, as $\mu$ decreases from $m_\tau$, $x$ increases 
from $x_\tau$  to a maximum value of 0.5625, which is achieved at a critical infrared 
momentum scale
\begin{equation}
\mu_c = m_\tau \left( \frac{9}{16x_\tau} \right)^{32/81} \exp \left[ \frac{32}{81} - \frac{2}{9x_\tau} \right]
\end{equation}
The domain of $x(\mu)$ is restricted to $\mu > \mu_c$, in which case $\mu_c$ represents an infrared-boundary 
of QCD at which the gauge couplant achieves its largest possible {\it but finite} value. The 
existence of an infrared boundary is no longer coupled with the arbitrary growth of the 
interaction coupling. 

By contrast, the $\beta$-function of eq. (5) has a 
zero at $x_{IRFP} = -8/(9R_2) + \sqrt{64/(81R_2^2)-1/R_2}$ ($R_2$ is negative), meaning 
that $x(\mu)$ increases to eventually level 
off at $x_{IRFP}$ as $\mu \rightarrow 0$. For this case, which corresponds to the existence of an 
infrared-stable fixed point (IRFP), {\it there is no infrared boundary} to 
gauge-theoretical QCD -- the domain of $x(\mu)$ includes all (positive) values of $\mu$. 

A point that needs to be made here is that if (1), (4), and (5) were to represent 
all-orders $\beta$-functions arising from legitimate but differing renormalization schemes,
\footnote{Equation (1) the $\beta$-function truncated after scheme-independent two-loop order
terms, has already been used to generate an apparently self-consistent renormalization scheme [2].}  
then the infrared behaviour of the couplant would itself be scheme-dependent.  Indeed, 
one could easily construct a $\beta$-function candidate consistent with (1) to two-loop order 
whose higher order terms are sufficiently large and negative [{\it e.g.} an arbitrarily large
negative $R_2$ in (5)] to ensure that an IRFP 
occurs at an arbitrarily {\it small} value for $x$, suggesting that QCD in such a scheme would 
remain a tractable perturbative gauge theory at arbitrarily large distances.  Since the 
infrared behaviour of QCD is empirically {\it known} to be confining at sufficiently large 
distances, such apparent scheme-dependence must be overstated. We therefore take the point 
of view here that differing but self-consistently realized renormalization schemes will 
necessarily lead to the same physical results -- in particular, to equivalent infrared 
dynamics.

A vector-like gauge theory that illustrates all this is that of $N = 1$ supersymmetric 
SU(3) Yang-Mills theory, supersymmetric gluodynamics characterised only by gluon and 
gluino fields.  Because of its supersymmetry, the all-orders $\beta$-function for this theory 
can be obtained algebraically by requiring that the anomaly multiplet within the theory 
respect the Adler-Bardeen theorem [3,4], or alternatively via instanton-calculus 
considerations [5]. Surprisingly, the same $\beta$-function arises from either approach, 
and even more surprisingly, this $\beta$-function, like (4), is characterised {\it by a pole}:   
\begin{equation}
\mu^2 \frac{dx}{d\mu^2} = -\frac{9x^2}{1-6x}, \; \; x \equiv g^2(\mu)/16\pi^2
\end{equation}
As noted in [6], such a pole necessarily implies a critical momentum scale $\mu_c$ for the 
asymptotically free phase of the theory that would constitute an infrared boundary to the 
region for which the theory is perturbative. The pole value of the couplant $[x(\mu_c) = 1/6]$ 
constitutes an infrared-attractive point terminating the evolution of the (real) gauge 
coupling constant in its asymptotically-free phase.  We henceforth denote (7) as the 
NSVZ $\beta$-function, after the authors of ref. 5.

The $\beta$-function for supersymmetric gluodynamics has also been calculated to three non-leading 
orders within the dimensional-reduction (DRED) renormalization scheme [7]:
\begin{equation}
\mu^2 \frac{dx}{d\mu^2} = -9x^2 [1+6x+63x^2+918x^3 + ... ]
\end{equation}
The higher order terms of the $\beta$-function (8) differ from those of the geometric 
series implicit in (7) once one gets beyond their equivalent leading and next-to-leading 
terms.  Both NSVZ and DRED schemes are presumably valid ones (although only the former 
upholds the Adler-Bardeen Theorem  to all orders). One can even obtain a perturbative 
road-map between couplants in the two schemes: if $z \equiv x^{NSVZ}$ and $y \equiv x^{DRED}$, 
then [4]
\begin{equation}
y=z\left[ 1+\sigma z + (27+6\sigma+\sigma^2) z^2 + (351+117\sigma+15\sigma^2+\sigma^3)z^3 + ... \right]
\end{equation}
where the constant $\sigma$ is {\it arbitrary} as a consequence of both 
(asymptotically-free) couplants having identical leading and next-to-leading terms within 
their respective $\beta$-function 
series.\footnote{An arbitrary constant in the relation between perturbative couplants 
in two different renormalization schemes of a given theory will occur provided the leading and next-to-leading terms of the $\beta$-function
series in both schemes are the same.  Such is the case, of course, for conventional QCD.}
Consequently, unless one specifies initial values for couplants in each scheme, 
one cannot say anything at all about the relative size of couplants in the two different 
schemes at a given momentum scale. If, for example, the all-order extension of the DRED 
series (8) exhibits a pole, such a pole need not occur at its $x = 1/6$ NSVZ location. 
Nevertheless, consistent infrared dynamics {\it does require} that a pole indeed occur in the DRED 
scheme. In other words, the all-orders extension of (8) should {\it} not exhibit the 
Landau-pole or IRFP behaviour seen to characterise of the $\beta$-function examples (1) and (5), 
respectively. Rather, the DRED scheme should the same infrared dynamics as are evident 
from (7): a finite infrared-attractive point $x(\mu_c)$ terminating the evolution of the 
asymptotically-free phase of the couplant, with $\mu_c$ serving as an infrared bound on the 
domain of $x(\mu)$.  

To get some insight into the infrared behaviour of DRED supersymmetric gluodynamics, the 
four known terms of the $\beta$-function series (8) have been utilised to construct Pad\'e 
approximants -- ratios of polynomials in $x$ whose power-series expansions reproduce 
these known terms. Pad\'e approximants are often employed not only to predict next 
order terms of a series, such as the $\beta$-function series for QCD [8], scalar field 
theory [8,9,10], and supersymmetric QCD [4,11], but also to explore whether infinite 
series with only a few known terms can be expected to have zeros {\it or poles} [12].
\footnote{Straightforward examples of this method for recovering the first positive pole
or zero of $\sec (x) \pm \tan(x)$ are presented in ref. [13].} In reference
[4], for example, it is demonstrated that every  Pad\'e approximant (except 
the truncated series itself) that reproduces the first four known terms of a series 
which differs infinitesimally from the geometric series $1/(1-|r|x)$ exhibits a pole 
that 
\begin{itemize}
\item[1)] differs infinitesimally from the true pole at $x =1/|r|$, and 
\item[2)] occurs prior 
to any positive spurious Pad\'e-approximant zeros.
\end{itemize}
In other words, the pole driven 
infrared dynamics of the NSVZ $\beta$-function (7) could have been predicted from Pad\'e 
approximants constructed from that series' first four terms.

Relevant Pad\'e approximants whose power series reproduce the four known terms of the 
DRED $\beta$-function (8) are [4]
\begin{equation}
\mu^2 \frac{dx}{d\mu^2} = -9x^2 \frac{(1-14x)}{(1-20x+57x^2)}
\end{equation}
\begin{equation}
\mu^2 \frac{dx}{d\mu^2} = -9x^2 \frac{(1-8.5714x-24.4286x^2)}{(1-14.5714x)}
\end{equation}
\begin{equation}
\mu^2 \frac{dx}{d\mu^2} = -9x^2 \frac{1}{(1-6x-27x^2-810x^3)}
\end{equation}
The first positive pole of (10) occurs at $x = 0.0604$ and precedes the zero at $x = 1/14$. 
Similarly, the pole of (11) at $x = 1/14.5714 = 0.0686$ precedes that approximant's first 
positive zero $(x = 0.0924)$, and the approximant (12) exhibits a first positive pole at 
$x = 0.0773$. Although (12) is constructed to have no zeros other than $x = 0$, there is no 
{\it a priori} reason that this approximant should exhibit a positive pole of comparable 
magnitude to those of (10) and (11).  These results (and further analysis of higher 
approximants presented in [4]) are clearly indicative of the same pole-driven infrared 
dynamics that characterise the NSVZ $\beta$-function (7) for the same $N = 1$ supersymmetric 
SU(3) Yang-Mills field theory. 

The techniques illustrated above are applicable to QCD itself. There is considerable 
controversy as to the infrared dynamics which characterise QCD, as well as the flavour 
dependence of such dynamics.  The idea that the QCD couplant freezes out to some 
effective $\alpha_s(\mu=0)$ at sufficiently low momentum scales appears to have both some 
theoretical justification [14] (even for the $n_f = 0$ case [15]) and as well as 
phenomenological utility [16]. However, such IRFP dynamics for $n_f \stackrel{<}{_\sim} 6$
are inconsistent with both theoretical studies based upon a $\beta$-function series truncated after its first 
two scheme independent terms [17] as well as with a lattice study indicative of an 
$n_f = 7$ threshold for IRFP dynamics [18].  

A Pad\'e-approximant approach to the infrared 
dynamics of QCD similar to the example presented above is formulated in detail in 
refs. [9], [13], and [19]. In this work, the known (and first unknown) terms of 
the QCD $\overline{MS}$ $\beta$-function series [20]
\begin{equation}
\mu^2 \frac{dx}{d\mu^2} = -\beta_0 x^2 S (x), \; \; \beta_0=11/12 - n_f / 6, \; \; 
x \equiv \alpha_s (\mu)/\pi
\end{equation}
\begin{eqnarray}
S(x) & = & 1+ \left[(51/8 - 19n_f / 24)/\beta_0 \right] x \nonumber\\
& + & \left[ (2857/2 - 5033 n_f/18+325n_f^2 / 54) /64\beta_0 \right]x^2 \nonumber\\
& + & \left[(114.23-27.134n_f+1.5824 n_f^2+5.8567 \cdot 10^{-3} n_f^3)/ \beta_0 \right]x^3 \nonumber\\
& + & R_4 x^4 + ...
\end{eqnarray}
are used to construct various Pad\'e-approximants whose leading positive zeros/poles 
are compared to see which occur first (or occur at all). We choose to use the 
$\overline{MS}$ renormalization shceme simply because the $\beta$-function series
in this scheme is known to higher order than in any other perturbative scheme -- moreover,
phenomenological QCD is overwhelmingly based upon $\overline{MS}$ calculations.
We denote by $S^{[N|M]}(x)$ 
the Pad\'e-approximant to $S(x)$ whose numerator and denominator are respectively 
degree-N and degree-M polynomials of the couplant $x$. If $N + M = 3$, the power-series 
expansion of the approximant $S^{[N|M]}(x)$ is of sufficiently high order to reproduce 
the known series terms in (14); {\it e.g.} for $n_f = 3$
\begin{equation}
S^{[2|1]}(x) = \frac{1-2.9169x - 3.8750 x^2}{1-4.6947 x}
\end{equation}
\begin{equation}
S^{[1|2]}(x) = \frac{1-8.1734x}{1-9.9511x + 13.220x^2}
\end{equation}

In both of these approximants, a positive pole precedes any positive zeros, indicative of 
the same pole-driven infrared dynamics known to characterise NSVZ supersymmetric 
gluodynamics. If one constructs such approximants for any choice of $n_f$ (as in ref. [13]), 
one finds for $n_f \leq 5$ that $S^{[2|1]}$ always has a positive pole preceding any positive zeros; 
moreover, the same statement applies to $S^{[1|2]}$ as well, provided $n_f \leq 6$. These results 
suggest that pole-driven infrared dynamics characterise QCD for up to five or six flavours.  
The occurrence of positive zeros that are not preceded by positive poles, ({\it i.e.}, 
$\beta$-function zeros corresponding to IRFP dynamics) does not characterise $S^{[2|1]}$ 
until $n_f \geq 7$, and does not characterise $S^{[1|2]}$ until $n_f \geq 9$, corroborating the 
existence of a flavour threshold of IRFP behaviour [17,18] as well as the absence of 
such behaviour below this threshold [21]. 

In ref. [13], these same qualitative conclusions are shown to be upheld by $S^{[N|M]}$  
for N+M =5, {\it i.e.}, for  Pad\'e-approximants to $S(x)$ whose power series are constructed 
to reproduce the four known terms of (14) and the {\it arbitrary} five loop term $R_4 x^4$. In 
such approximants, each numerator and denominator polynomial coefficient of powers 
of $x$ is itself linear in the parameter $R_4$ -- {\it e.g.} for $n_f = 3$
\begin{equation}
S^{[2|2]} (x) = \frac{1+(7.1946 - 0.10261 R_4)x + (-11.329+0.075644 R_4) x^2}
{1+(5.4168-0.10261 R_4)x + (-25.430 + 0.25806 R_4)x^2}
\end{equation}
{\it Regardless of the value $R_4$ takes}, one finds that positive zeros which precede every 
positive pole do not occur in the approximants $S^{[1|3]}$, $S^{[2|2]}$ , or $S^{[3|1]}$ until  
$n_f \geq$ 9, 7, and 6, respectively, consistent with the non-occurrence of IRFP dynamics 
below these threshold values of $n_f$. Moreover, for arbitrary $R_4$ a positive pole which 
precedes every positive zero {\it does} occur in the approximants $S^{[1|3]}$ and $S^{[2|2]}$, 
provided $n_f \leq 5$. The approximant $S^{[3|1]}$ has a single positive pole only if $R_4 > 0$, 
a consequence of having a denominator linear in $x$, but this pole is found to precede 
any numerator zeros in the approximant for all positive values of $R_4$ as long as $n_f \leq 7$. 

These results are clearly indicative of the occurrence of pole-driven infrared dynamics 
for QCD once heavy flavours are decoupled. Similar pole-driven infrared dynamics are also shown
in [13] to characterise QCD in the $N_c \rightarrow \infty$ 't Hooft limit. 
Moreover, very recent work [18] has demonstrated that different approximants to the 
$N_c = 3, \; n_f = 3 \; \overline{MS}$ QCD $\beta$-function 
exhibit surprising consistency in their predictions of infrared boundary coordinates 
$(\mu_c, x(\mu_c))$ associated with pole-driven infrared dynamics. The implications of such 
dynamics, in particular the possibility of having both a strong and an asymptotically-free 
phase of QCD with common infrared properties [6], have only begun to be explored.

\newpage

\noindent{\large\bf Acknowledgements}\\

I am grateful to my research collaborators F. A. Chishtie, V. A. Miransky, and 
T. G. Steele, who coauthored of much of the research described above.  I am 
also indebted to my deceased research collaborator Mark Samuel, who  pioneered 
the application of Pad\'e approximants to perturbative quantum field theory, and 
to Roger Migneron, whose final research paper [9] is the first published work in 
which Pad\'e-approximants are applied to the infrared structure of QCD.

\bigskip

\noindent{\large\bf References}\\
\begin{itemize}

\item[1.]D. E. Groom et al. [Particle Data Group], Eur. Phys. J. C {\bf 15}, 1 (2000).

\item[2.]G. 't Hooft, in Recent Developments in Gauge Theories, Vol. 59 of NATO 
Advanced Study Institute Series B: Physics, edited by G. 't Hooft et al. (Plenum, N.Y., 1980).

\item[3.]D. R. T. Jones, Phys. Lett. B {\bf 123}, 45 (1983).

\item[4.]V. Elias, J. Phys. G {\bf 27}, 217 (2001).

\item[5.]V. Novikov, M. Shifman, A. Vainshtein, and V. Zakharov, Nucl. Phys. B 229, 381 (1983).

\item[6.]I. I. Kogan and M. Shifman, Phys. Rev. Lett. {\bf 75}, 2085 (1995).

\item[7.]L. N. Avdeev, G. A. Chochia, and A. A. Vladimirov, Phys. Lett. B {\bf 105}, 272 (1981); 
I. Jack, D. R. T. Jones, and A. Pickering, Phys. Lett. B 435, 61 (1998).

\item[8.]J. Ellis, M. Karliner, and M. A. Samuel, Phys. Lett. B {\bf 400}, 176 (1997).

\item[9.]V. Elias, T. G. Steele, F. Chishtie, R. Migneron, and K. Sprague, Phys. Rev. D 
{\bf 58}, 116007 (1998).

\item[10.]F. Chishtie, V. Elias, and T. G. Steele, Phys. Lett. B {\bf 466}, 266 (1999); 
F. A. Chishtie and V. Elias, Phys. Lett. B {\bf 499}, 270 (2001).

\item[11.]I. Jack, D. R. T. Jones, and M. A. Samuel, Phys. Lett. B {\bf 407}, 143 (1997); 
J. Ellis, I. Jack, D. R. T. Jones, M. Karliner, and M. A. Samuel, Phys. Rev. D {\bf 57}, 2665 (1998).

\item[12.]G. Baker and P. Graves-Morris, Padé Approximants [Vol. 13 of Encyclopedia of 
Mathematics and its Applications] (Addison-Wesley, Reading, MA, 1981) pp. 48-57.

\item[13.]F. A. Chishtie, V. Elias, V. A. Miransky, and T. G. Steele, Prog. Theor. Phys. 
{\bf 104}, 603 (2000).

\item[14.]A. C. Mattingly and P. M. Stevenson, Phys. Rev. Lett. {\bf 69}, 1320 (1992); 
P. M. Stevenson, Phys. Lett. B {\bf 331}, 187 (1994).

\item[15.]A. C. Mattingly and P. M. Stevenson, Phys. Rev. D {\bf 49}, 437 (1994).

\item[16.]Yu. L. Dokshitzer, in Proceedings of the 29th International Conference in 
High Energy Physics, A. Astbury, D. Axen, and J. Robinson, eds. (World Scientific, 
Singapore, 1999) pp. 305-324.

\item[17.]T. Banks and A. Zaks, Nucl. Phys. B {\bf 196}, 189 (1982); 
T. Appelquist, J. Terning, and L. C. R. Wijewardhana, Phys. Rev. Lett. {\bf 77}, 1214 (1996); 
V. A. Miransky and K. Yamawaki, Phys. Rev. D55, 5051 and (Err.) D {\bf 56}, 3768 (1997).

\item[18.]Y. Iwasaki, K. Kanaya, S. Sakai, and T. Yoshié, Phys. Rev. Lett. {\bf 69}, 21 (1992).

\item[19.]F. A. Chishtie, V. Elias, and T. G. Steele, Phys. Lett. B {\bf 514}, 279 (2001).

\item[20.]T. van Ritbergen, J. A. M. Vermaseren, and S. A. Larin, Phys. Lett. B {\bf 405}, 
323 (1997).

\item[21.]E. Gardi, G. Grunberg, and M. Karliner, JHEP 9807, 007 (1998).

\end{itemize}
\end{document}